\documentclass[preprint]{revtex4}
\usepackage{amsmath,amssymb,amsthm,graphicx,slashed}
\newtheorem{thm}{Theorem}%[section]
\newtheorem{corl}[thm]{Corollary}
\newtheorem{lma}[thm]{Lemma}
\newtheorem{prop}[thm]{Proposition}

\def\A{\mathcal{A}}

\def\B{\mathcal{B}}
\def\bar{\overline}
\def\C{\mathbb{C}}

\def\E{\mathcal{E}}
\def\ev{\textup{ev}}
\DeclareMathOperator{\End}{End}
\def\bH{\mathbb{H}}
\def\H{\mathcal{H}}
\def\pert{{\rm Pert}}
\def\R{\mathbb{R}}
\DeclareMathOperator{\tr}{Tr}
\def\U{\mathcal{U}}
\def\qqq{\,,\quad~\forall}
\def\dd{{\rm Dom}}
\newcommand{\ie}{{\it i.e.\/}\ }
\begin{document}

\title[Inner Fluctuations]{Inner Fluctuations in Noncommutative Geometry \\without the first order condition}
\author{Ali H. Chamseddine$^{1,3}$, Alain Connes$^{2,3,4}$ and Walter D. van Suijlekom$^{5}$}
\email{chams@aub.edu.lb, alain@connes.org, waltervs@math.ru.nl}
\affiliation{$^{1}$Physics Department, American University of Beirut, Lebanon}
\affiliation{$^{2}$College de France, 3 rue Ulm, F75005, Paris, France}
\affiliation{$^{3}$I.H.E.S. F-91440 Bures-sur-Yvette, France}
\affiliation{$^{4}$Department of Mathematics, The Ohio State University, Columbus OH 43210 USA}
\affiliation{$^{5}$Institute for Mathematics, Astrophysics and Particle Physics,
Radboud University Nijmegen, Heyendaalseweg 135, 6525 AJ Nijmegen, The Netherlands}


\begin{abstract}
We extend inner fluctuations to spectral triples that do not fulfill the first-order condition. This involves the addition of a quadratic term to the usual linear terms. We find a semi-group of inner fluctuations, which only depends on the involutive algebra $\A$ and which extends the unitary group of $\A$. This has a key application in noncommutative spectral models beyond the Standard Model, of which we consider here a toy model. 
\end{abstract}

\maketitle

\section{Introduction}

Noncommutative geometry provides a new paradigm of geometric space, coming from and expressed in the language of quantum mechanics, \ie that of operators in Hilbert space. The space itself is encoded by its algebra $\A$ of coordinates which is concretely represented as operators in a Hilbert space $\H$. The geometry of the space is encoded by its inverse line element which is also an operator $D$ acting in the same Hilbert space. Ordinary geometric spaces, \ie Riemannian manifolds $(X, g_{\mu\nu})$, fit in this framework using the Hilbert space $\H=L^2(X,S)$ of spinors, the action of the algebra of functions on $X$ by multiplication, and the Dirac operator $D$.

While this appears at first as a reformulation of geometric notions in an algebraic language, one gets an immediate reward which is a complete list of gravitational observables, \ie of diffeomorphism invariant quantities of the given geometry $(X, g_{\mu\nu})$. They are given by the spectrum of the operator $D$ and the relative position (generalized angle) of the two algebras of operators given by $\A$ on one hand and the algebra of functions of $D$ on the other.
Moreover the Einstein-Hilbert action itself is easily expressed as a spectral invariant of the operator $D$.

But another key virtue of the new paradigm is that it does not require the commutativity of the algebra $\A$. The gauge sector of the Standard Model is non-abelian and the possibility to extend geometry to the noncommutative case allows one to consider for instance the algebra $\A$ of matrices $M_n$ of functions on an ordinary manifold $X$. What one finds is that the theory of ``pure gravity'' on such a slightly noncommutative space, gives ordinary gravity on $X$ minimally coupled with pure Yang-Mills theory for the gauge group $SU(n)$.

This is a good indication but such simplistic gauge models are still quite far from the intricacies of the Standard Model minimally coupled with gravity, and for a long time the leitmotif of the bottom-up approach of noncommutative geometry has been to understand where the above fundamental physical model fits in the new paradigm of geometry. This culminated in \cite{mc2} with a noncommutative geometric formulation of the Standard Model, including the full Higgs sector and see-saw mechanism, unified with gravity.

From a mathematical perspective, the spectral Standard Model appears naturally in the classification of irreducible finite geometries of $KO$-dimension $6$ (modulo $8$) performed in \cite{CC07b}. The breaking of the natural algebra $\bH \oplus \bH \oplus M_4(\C)$ which results from that classification to the subalgebra $\C \oplus \bH \oplus M_3(\C)$ corresponding to the Standard Model was effected using the requirement of the first order condition on the Dirac operator.

In this paper we analyze the first order condition and the above breaking much further. 
We shall illustrate our analysis by a simplified case here and treat the case of the full standard model in a forthcoming paper. 
The origin of the first order condition and its name come from the algebraic encoding of the fact that the Dirac operator over an ordinary geometric space is a differential operator of order one. It is not a derivation of the algebra of functions $\A$ into itself but a derivation of the algebra $\A$ into the commutant of $\hat \A=J\A J^{-1}$ where the antilinear isometry $J:\H\to \H$ takes its origin in the work of Tomita, and fulfills the commutativity condition
\begin{equation}\label{com}
[a,Jb J^{-1}]=0 \qqq a,b \in \A.
\end{equation}
Thus the first order condition is
\begin{equation}\label{first}
    [[D,a],Jb J^{-1}]=0 \qqq a,b \in \A.
\end{equation}
The operator $J$ is a simultaneous incarnation of Tomita's anti-isomorphism operator, of the charge conjugation operator and of the nuance between a $KO$-homology cycle and a $K$-homology cycle.

So far, the notion of inner fluctuations of the noncommutative geometry was developed under the
requirement of the first order condition. The fluctuated  metrics are of the form
\begin{equation}\label{innerfluc}
D'=D+A+\epsilon JAJ^{-1}, \ \ A=\sum a_j[D,b_j]
\end{equation}
where $\epsilon\in\{ \pm 1\}$ is such that $JDJ^{-1} =\epsilon D$ and $\omega=\sum a_j\delta(b_j)\in \Omega^1(\A)$ is a self-adjoint universal one form (we denote by $\delta$ rather than $d$ the universal differential). These fluctuations are the counterpart for the metric of the inner fluctuations of the automorphisms of the algebra. The inner automorphisms are the automorphisms $\alpha$ of $\A$ of the form $\alpha(x)=uxu^*$ (for a unitary $u\in \A$) and they form a normal subgroup of the group of automorphisms which plays the same role as the group of gauge transformations as a subgroup of the symmetry group of the Lagrangian of gravity coupled with matter. The first order condition
is essential in order to guarantee the gauge invariance of the inner fluctuations under the action of the gauge group given by the unitaries $U=uJuJ^{-1}$ for any unitary $u\in \A$.

Our point of departure for this paper is that one can extend \eqref{innerfluc} to the general case, \ie without assuming the order one condition. It suffices to add a quadratic term which only depends upon the universal $1$-form $\omega\in \Omega^1(\A)$ to the formula \eqref{innerfluc} and one restores in this way,
\begin{itemize}
  \item The gauge invariance under the unitaries $U=uJuJ^{-1}$
  \item The fact that inner fluctuations are transitive, \ie that inner fluctuations
  of inner fluctuations are themselves inner fluctuations.
\end{itemize}
We show moreover that the resulting inner fluctuations come from the action on operators in Hilbert space of a semi-group $\pert(\A)$ of {\em inner perturbations} which only depends on the involutive algebra $\A$ and extends the unitary group of $\A$.
This opens up two areas of investigation, the first is mathematical and the second is directly related to particle physics and model building:
\begin{enumerate}
  \item Investigate the inner fluctuations for noncommutative spaces such as quantum groups
  and quantum spheres.
  \item Compute the spectral action and inner fluctuations for the model involving the full symmetry algebra $\mathbb H\oplus \mathbb H\oplus M_4(\C)$ before the breaking to the Standard Model algebra.
\end{enumerate}
For the first point we refer to \cite{diracs2q,diracsu2,indexsu2,diracalls2q} for natural examples of spectral triples not fulfilling the first order condition. The second point is the contents of a separate paper \cite{ncgps}. In the present paper we will treat a simpler toy model, involving the algebra $\C \oplus \C \oplus M_2(\C)$. It serves as an illustration of the generalized form of the inner fluctuations, the spontaneous symmetry breaking mechanism appearing in the spectral action, and as a preparation for the full model based on $\bH \oplus \bH \oplus M_4(\C)$. In the aforementioned classification of irreducible finite geometries of KO-dimension 6 (modulo 8) of \cite{CC07b}, this toy model corresponds to the case $k=2$ (whereas the full model corresponds to $k=4$). We note that the case $k=8$ was discussed recently in a slightly different context in \cite{DLM13}.

\section{Acknowledgments} AHC is supported in part by the National Science
Foundation under Grant No. Phys-0854779 and Phys-1202671. WDvS thanks IH\'ES for hospitality during a visit from January-March 2013.

\section{First-order condition and inner fluctuations}
We generalize inner fluctuations to real spectral triples that fail on the first-order condition. In this case, the usual prescription \cite{C96} does not apply, since the operator $D+A+\epsilon JAJ^{-1}$ with $A \in \Omega^1_D(\A)$ does not behave well with respect to the action of the gauge group $\U(\A)$. In fact, one would require that conjugation of the fluctuated Dirac operator by the unitary operator $U:=u Ju J^{-1}$ for $u \in \U(\A)$ can be implemented by a usual type of gauge transformation $A \mapsto A^u = u[D,u^*]+ uAu^*$ so that
$$
D+A+\epsilon JAJ^{-1} \mapsto
U (D+A+\epsilon JAJ^{-1}) U^* \equiv D+A^u +\epsilon  JA^uJ^{-1}
$$
However, the simple argument only works if $[JuJ^{-1},A] = 0$ for all $A\in \Omega^1_D(\A)$ and $u \in \U(\A)$, that is, if the first-order condition is satisfied.

\subsection{Morita equivalence and spectral triples}
We start with the following general result on Morita equivalence for spectral triples $(\A,\H,D;J)$ that possibly do not satisfy the first-order condition. First, we introduce some notation. If $\E$ is a finite-projective right $\A$-module, equipped with a universal connection $\nabla: \E \to \E \otimes_\A \Omega^1(\A)$, we introduce the following linear operator on $\E \otimes_\A \H$
\begin{align*}
\left( 1 \otimes_\nabla D \right)(v \otimes \xi) &:= \nabla_D(v) \xi + v \otimes D \xi; \qquad (v \in \E, \xi \in \dd(D)\subset\H),
\end{align*}
where $\nabla_D(v)$ indicates that universal one-forms $ a\delta(b) \in \Omega^1(\A)$ are represented as $a[D,b]\in \Omega^1_D(\A)$ using the Dirac operator $D$.

Subsequently, we can introduce a linear operator $(1 \otimes_\nabla D) \otimes_{\bar \nabla} 1$ on $\E \otimes_\A \H \otimes_\A \bar\E$ using the induced conjugate {\em left} connection $\bar\nabla: \bar\E \to \Omega^1(\A) \otimes_\A \bar \E$. Explicitly, it is given by
$$
\left((1 \otimes_\nabla D) \otimes_{\bar \nabla} 1\right)(v_1 \otimes \xi \otimes \bar v_2)
= \nabla_D(v_1) (\xi) \otimes \bar v_2 + v_1 \otimes D\xi \otimes \bar v_2
+ (v_1 \otimes \xi)\bar\nabla_{1 \otimes_\nabla D} (\bar v_2).
$$
for $v_1,v_2 \in \E, \xi \in \dd(D)\subset \H$.

Similarly, we can define $1 \otimes_\nabla (D \otimes_{\bar \nabla} 1)$ as an operator on $\E \otimes_\A \H \otimes_\A \bar\E$.

\begin{prop}\label{innfluc}
Let $(\A,\H,D;J)$ be a real spectral triple, possibly not fulfilling the first-order condition. Let $\E$ be a finite-projective right $\A$-module, equipped with a {\em universal} connection $\nabla :  \E \to \E \otimes_{\A } \Omega^1(\A)$. Then
\begin{equation}\label{assoc}
   (1 \otimes_\nabla D) \otimes_{\bar \nabla} 1=1 \otimes_\nabla (D \otimes_{\bar \nabla} 1)
\end{equation}
 Moreover the triple
 $(\End_{\A}(\E), \E \otimes_{\A} \H \otimes_\A \bar\E, D';J')$ is a real spectral triple where
$D'= (1 \otimes_\nabla D) \otimes_{\bar \nabla} 1$
and the real structure is given by
\begin{equation*}
J'(v_1 \otimes \xi \otimes\bar v_2) = (v_2 \otimes J\xi \otimes\bar v_1); \qquad (v_1,v_2 \in \E, \xi \in \H).
\end{equation*}

%where a universal one-form of the form $a \delta(b)$ ($a,b\in \A$) entering in $\nabla(e_1)$ acts on $\H \otimes_\A \bar \E$ as the operator $a[D\otimes_{\bar\nabla} 1,b]$, and the universal one-form $a \delta(b)$ entering in $\bar\nabla(\bar e_2)$ as $J (a[1 \otimes_\nabla D,b])^* J^{-1}$.
\end{prop}
\proof Since the module $\E$ is finite and projective, one can find an integer $n$ and an idempotent $e\in M_n(\A)$ such that $\E$ is isomorphic to the right module $e\A^n$. One then has \begin{equation}\label{tens}
    \E \otimes_{\A } \Omega^1(\A)\sim e\left( \Omega^1(\A)\right)^n, \ \
    \E \otimes_{\A } \H\sim e\H^n
\end{equation}
In order to prove \eqref{assoc} we first assume that the connection $\nabla :  \E \to \E \otimes_{\A } \Omega^1(\A)$ is the Grassmannian connection given by
\begin{equation}\label{grass}
    \nabla((v_j)):=e((\delta (v_j)))\in e\left( \Omega^1(\A)\right)^n\qqq v_j\in \A,\ 1\leq j\leq n.
\end{equation}
For such a connection one gets, for $\xi_i\in\dd(D)\subset \H,\ 1\leq i\leq n$ such that $e(\xi_i)=(\xi_i)$,
\begin{equation}\label{grassconnect}
    (1 \otimes_\nabla D)((\xi_i))=e((D\xi_i))\in  e\H^n
   \end{equation}
Indeed one has, in terms of the matrix components $e_{ij}$ of $e\in M_n(\A)$,
\begin{equation*}
   (1 \otimes_\nabla D)(e_{ij}\otimes \xi_j)=
   \nabla_D(e_{ij}) \xi_j + e_{ij} \otimes D \xi_j\qqq j
\end{equation*}
but since $e^2=e$ one gets that, in  $\Omega^1(\A)$ and for each $i,\ell$
\begin{equation*}
   \sum_{j,k} e_{ij}\delta(e_{jk})e_{k\ell}=0
\end{equation*}
as one shows using $\delta(e)=\delta(e^2)=\delta(e)e+e\delta(e)$ and $e\delta(e)e=0$ where $\delta$ is applied componentwise to the matrix $e_{ij}$. It thus follows that the terms in $\nabla_D(e_{ij})\xi_j$ sum to $0$
and one gets \eqref{grassconnect}.

Next one has
\begin{equation*}
    \E \otimes_\A \H \otimes_\A \bar\E\sim \pi(e)\hat\pi(e)M_n(\H)
\end{equation*}
where the representations $\pi$ and $\hat\pi$ of the real algebra $M_n(\A)$ are given by
\begin{equation*}
   ( \pi(a)\xi)_{ij}=\sum a_{ik}\xi_{kj}, \ \ (\hat\pi(a)\xi)_{ij}=\sum \hat a_{jk}\xi_{ik}
\end{equation*}
and we introduce, for operators in $\H$, the notations
\begin{equation}\label{bar}
   \hat T=JTJ^{-1},  \   \    T^\circ =JT^*J^{-1}.
\end{equation}
Note that $T\mapsto \hat T$  defines an antilinear automorphism on operators (thus $\widehat{AB}=\hat A\hat B$), not to be confused with the linear antiautomorphism 
$T\mapsto T^\circ 
$ which reverses the order of the terms in a product.
The two representations $\pi$ and $\hat\pi$  commute, and one gets, with $\tilde D=1_{M_n(\C)}\otimes D$ acting in $M_n(\H)$, that
\begin{equation*}
   (1 \otimes_\nabla D) \otimes_{\bar \nabla} 1=\hat\pi(e)\left(\pi(e)\tilde D\right)
   =\hat\pi(e)\pi(e)\tilde D=\pi(e)\hat\pi(e)\tilde D=1 \otimes_\nabla (D \otimes_{\bar \nabla} 1)
\end{equation*}
To pass from this particular connection $\nabla_0$ to the general case, one expresses an arbitrary connection as $\nabla=\nabla_0+ eAe$ where $A=(A_{ij})$ is a matrix of one forms
$A_{ij}\in \Omega^1(\A)$. The computation is then the same as in the case of the trivial right module $\E=\A$ which we shall do in details in \S \ref{sectinnerf} below.\endproof

\begin{corl}
If $(\A,\H,D;J)$ satisfies the first-order condition, then so does  $(\End_{\A}(\E), \E \otimes_{\A} \H \otimes_\A \bar\E, D';J')$ and in that case the above inner fluctuation reduces to the usual one, given in terms of a connection $\nabla: \E \to \E \otimes_\A \Omega^1_D(\A)$ ({\it i.e.} representing all universal connections using $\delta \mapsto [D,\cdot]$).
\end{corl}

\subsection{Special case $\E = \A$ and inner fluctuations}\label{sectinnerf}
As a special case we take $\E = \A$ and $\nabla= \delta + A$ where $A \in \Omega^1(\A)$ is a self-adjoint, {\em universal} one-form
\begin{equation}\label{form}
A = \sum_j a_j \delta(b_j); \qquad (a_j,b_j \in \A).
\end{equation}
Under the respective identifications $\H=\A \otimes_\A \H$ and $\H=\H \otimes_\A \A$, we have
\begin{align*}
1 \otimes_\nabla D &\simeq D+\sum_j a_j[D,b_j],\\
D \otimes_{\bar \nabla} 1 &\simeq  D+\sum_j \hat a_j[D,\hat b_j]  .
\end{align*}
This then gives rise to the following Dirac operator
\begin{align}\label{opDprime}
D' &= D+\sum_j a_j  [D, b_j ] + \sum_j \hat a_j[D,\hat b_j] +
 \sum_j \hat a_j [A_{(1)}, \hat b_j] \nonumber \\
&=: D+A_{(1)}+\tilde A_{(1)} + A_{(2)}
\end{align}
where we have defined
\begin{align*}
A_{(1)} &:= \sum_j a_j [D,b_j];\\
\tilde A_{(1)} &:= \sum_j \hat a_j [D,\hat b_j];\\
A_{(2)} &:= \sum_j \hat a_j [ A_{(1)}, \hat b_j] \\&= \sum_{j,k} \hat a_j a_k [ [D,b_k], \hat b_j]
\end{align*}
The commutant property \eqref{com} shows that
\begin{equation*}
   \sum_j \hat a_j [A_{(1)}, \hat b_j]= \sum_{j,k} \hat a_j a_k [ [D,b_k], \hat b_j]
   = \sum_{j,k} a_k \hat a_j  [ [D, \hat b_j],b_k]
   =\sum_k a_k[\tilde A_{(1)},b_k]
\end{equation*}
which checks \eqref{assoc}.
 Note  that, with $\epsilon= \pm 1$ such that $JDJ^{-1}=\epsilon D$ one has
\begin{equation*}
\tilde A_{(1)}=\epsilon J A_{(1)}J^{-1},\ \ A_{(2)} =\epsilon J A_{(2)} J^{-1}
\end{equation*}
%$$
%A_{(2)} = \sum_{j,k} a_j [ Ja_kJ^{-1} [D,J b_kJ^{-1}], b_j]
%$$
which follows from the commutant property \eqref{com}.

It is clear from these formulas that $A_{(2)}$ vanishes if $(\A,\H,D;J)$ satisfies the first-order condition, thus reducing to the usual formulation of inner fluctuations.

We will interpret the terms $A_{(2)}$ as non-linear corrections to the {\em first-order}, linear inner fluctuations $A_{(1)}$ of $(\A,\H,D;J)$. It is clear that the first order condition is equivalent to the linearity of the map from $1$-forms to fluctuations.
In fact, gauge transformations act on $D'$ as:
$$
D'  \mapsto U D'  U^*
$$
with $U = u JuJ^{-1}$ and $u \in \U(\A)$. By construction, it is implemented by the gauge transformation
$$
A \mapsto u A u^* + u \delta(u^*)
$$
in the universal differential calculus. In particular, this implies that
$$
A_{(1)} \mapsto u A_{(1)} u^* + u [D,u^*] \in \Omega^1_D(\A)
$$
so the first-order inner fluctuations transform as usual. For the term $A_{(2)}$ we compute that a gauge transformation acts as
$$
A_{(2)} \mapsto JuJ^{-1} A_{(2)} J u^* J^{-1} + JuJ^{-1} [u[D,u^*],Ju^* J^{-1}]
$$
where the $A_{(2)}$ on the right-hand-side is expressed using the gauge transformed $A_{(1)}$. This non-linear gauge transformation for $A_{(2)}$ confirms our interpretation of $A_{(2)}$ as the non-linear contribution to the inner fluctuations.

Let us do the direct check that the gauge transformations operate in the correct manner thanks to the quadratic correction term $A_{(2)}$. We shall understand this direct computation in a more conceptual manner in \S \ref{sectsemigroup}.
\begin{lma} Let $A\in \Omega^1(\A)$ be a universal one form as in \eqref{form}, and $D'=D(A)$  be given by \eqref{opDprime}. Let $u \in \U(\A)$ and $U = u JuJ^{-1}$. Then one has
\begin{equation}\label{gauge}
    UD(A)U^*=D(\gamma_u(A)), \ \ \ \gamma_u(A)=u\delta(u^*)+uAu^*\in \Omega^1(\A)
\end{equation}
\end{lma}
\proof  Let $A=\sum_1^n a_j \delta(b_j)\in \Omega^1(\A)$, one has
\begin{equation*}
   \gamma_u(A)=u(1-\sum_1^n a_jb_j)\delta(u^*)+\sum_1^n ua_j\delta(b_ju^*)=\sum_0^n a'_j \delta(b'_j)
\end{equation*}
where $a'_0=u(1-\sum_1^n a_jb_j)$ and $b'_0=u^*$, while $a'_j=ua_j$ and $b'_j=b_ju^*$ for $j>0$.
What matters is the following, valid for any inclusion $\A\subset \B$, and $T\in \B$
\begin{equation}\label{comgauge}
    \sum_0^n a'_j [T,b'_j]=u[T,u^*]+u\left(\sum_1^n a_j[T,b_j]\right) u^*
\end{equation}
We use the notation \eqref{bar} for any operator in $\H$.
With this notation we have
\begin{align*}
A_{(1)} &:= \sum_j a_j [D,b_j];\\
A_{(2)} &:= \sum_j \hat a_j [ A_{(1)}, \hat b_j ] \\&= \sum_{j,k} \hat a_j  [ a_k [D,b_k], \hat b_j]
\end{align*}
We now apply these formulas using $\gamma_u(A)=\sum_0^n a'_j \delta(b'_j)$ and obtain using \eqref{comgauge},
\begin{equation}\label{gauge1}
    A'_{(1)}=u[D,u^*]+u\left(\sum_1^n a_j[D,b_j]\right) u^*=u[D,u^*]+u A_{(1)}u^*
\end{equation}
and
\begin{equation}\label{gaugeA2}
    A'_{(2)}=\sum_j \hat a'_j [ A'_{(1)}, \hat b'_j ]=
    \hat u [A'_{(1)}, \hat u^*]+\hat u\left(\sum_j \hat a_j [ A'_{(1)}, \hat b_j ]\right)\hat u^*
\end{equation}
So, using \eqref{gauge1}, we get (assuming to simplify that $\epsilon=1$ so $\hat D=D$)
\begin{equation*}
   \sum_j \hat a_j [ A'_{(1)}, \hat b_j ]=\sum_j \hat a_j [u[D,u^*], \hat b_j ]+\sum_j \hat a_j [ uA_{(1)}u^*, \hat b_j ]
\end{equation*}
and the commutation of the $\hat x$ with the $y$, for $x,y\in \A$ gives
\begin{equation*}
   \sum_j \hat a_j [ uA_{(1)}u^*, \hat b_j ]=u\left(\sum_j \hat a_j [ A_{(1)}, \hat b_j ]\right)u^*=u A_{(2)}u^*
\end{equation*}
and using $u[D,u^*]=uDu^*-D$,
\begin{equation*}
 \sum_j \hat a_j [u[D,u^*], \hat b_j ]=u\left(\sum_j \hat a_j [D, \hat b_j ]    \right)u^*-\sum_j \hat a_j [D, \hat b_j ]=u\hat A_{(1)}u^*- \hat A_{(1)}
\end{equation*}
so that we get:
\begin{equation}\label{gauge2}
    \hat u\left(\sum_j \hat a_j [ A'_{(1)}, \hat b_j ]\right)\hat u^*=
 \hat u   u \hat A_{(1)}u^*\hat u^*
 -\hat u\hat A_{(1)}\hat u^*\\+
  \hat u   u A_{(2)}u^*\hat u^*
\end{equation}
Next one has
\begin{equation*}
    \hat u [A'_{(1)}, \hat u^*]=\hat u [u[D,u^*], \hat u^*]
    + \hat u [uA_{(1)}u^*, \hat u^*]=\hat u [u[D,u^*], \hat u^*]
    + \hat u   u  A_{(1)}u^*\hat u^*-uA_{(1)}u^*
\end{equation*}
so that, using \eqref{gaugeA2} we obtain
\begin{equation}\label{gaugeA2new}
    A'_{(2)}=\hat u [u[D,u^*], \hat u^*]
    + U  A_{(1)}U^*-uA_{(1)}u^*+U \hat A_{(1)}U^*
 -\hat u\hat A_{(1)}\hat u^*+
  U A_{(2)}U^*
\end{equation}
We then obtain
\begin{equation*}
    A'_{(1)}+\hat A'_{(1)}+ A'_{(2)}=u[D,u^*]+\hat u[D,\hat  u^*]+\hat u [u[D,u^*], \hat u^*]+ U  A_{(1)}U^*+U \hat A_{(1)}U^*
+  U A_{(2)}U^*
\end{equation*}
and the result follows using
\begin{equation*}
 UDU^*=   D+u[D,u^*]+\hat u[D,\hat  u^*]+\hat u [u[D,u^*], \hat u^*].
\end{equation*}
\endproof

\subsection{The semigroup of inner perturbations}\label{sectsemigroup}
We show that inner fluctuations come from the action on operators in Hilbert space of a semi-group $\pert(\A)$ of {\em inner perturbations} which only depends on the involutive algebra $\A$ and extends the unitary group of $\A$. This covers both cases of ordinary spectral triples and real spectral triples (\ie those which are equipped with the operator $J$). In the latter case one simply uses the natural homomorphism of semi-groups $\mu:\pert(\A)\to \pert(\A\otimes \hat \A)$ given by $\mu(A)=A\otimes \hat A$. This implies in particular that inner fluctuations of inner fluctuations are still inner fluctuations and that the corresponding algebraic rules are unchanged by passing from ordinary spectral triples  to real spectral triples.

We  first show that the formulas of the previous sections can be greatly simplified by representing the universal $1$-forms as follows, where $\A^{\rm op}$ denotes the opposite algebra of  $\A$ and $x\mapsto x^{\rm op}$ the canonical antiisomorphism 
$\A\mapsto \A^{\rm op}$,
\begin{lma}
$(i)$~The following map $\eta$ is a surjection
\begin{equation*}
 \eta: \{\sum a_j\otimes b_j^{\rm op}\in  \A\otimes \A^{\rm op}\mid \sum a_jb_j=1\}
 \to \Omega^1(\A),\ \ \eta(\sum a_j\otimes b_j^{\rm op})=\sum a_j\delta( b_j).
\end{equation*}
$(ii)$~One has
\begin{equation*}
    \eta\left(\sum b_j^*\otimes a_j^{*\rm op}\right)=\left(\eta\left(\sum a_j\otimes b_j^{\rm op}\right)\right)^*
\end{equation*}
$(iii)$~One has, for any unitary $u\in \A$,
\begin{equation*}
    \eta\left(\sum u a_j\otimes (b_ju^*)^{\rm op}\right)=\gamma_u\left(\eta\left(\sum a_j\otimes b_j^{\rm op}\right)\right)
\end{equation*}
where $\gamma_u$ is the gauge transformation of potentials.
\end{lma}
\proof $(i)$~Let us start from an element $\omega=\sum x_i\delta(y_i)\in \Omega^1(\A)$. Then since $\delta(1)=0$ it is the same as
\begin{equation*}
   (1-\sum x_i y_i)\delta(1)+ \sum x_i\delta(y_i)
\end{equation*}
and one checks that the normalization condition is now fulfilled.

$(ii)$~The normalization condition is fulfilled by $\sum b_j^*\otimes a_j^{*\rm op}$ since $\sum b_j^*a_j^*=(\sum a_jb_j)^*$. Thus one gets the equality  using $\delta(x)^*=-\delta(x^*)$ and
\begin{equation*}
    \sum b_j^*\delta( a_j^*)=-\left(\sum \delta(a_j)b_j\right)^*=
    \left(\sum a_j\delta(b_j)\right)^*
\end{equation*}

$(iii)$~The normalization condition is fulfilled by $\sum ua_j\otimes (b_ju^*)^{*\rm op}$ since
$\sum ua_j b_ju^*=1$. Moreover one has, using $\delta (b_ju^*)=\delta(b_j)u^*+b_j\delta(u^*)$
\begin{equation*}
   \sum ua_j\delta (b_ju^*)=u\left(\sum a_j\delta(b_j)\right) u^*+u\delta(u^*)
\end{equation*}
\endproof

\begin{prop}\label{propinn}
$(i)$~Let $A=\sum a_j\otimes b_j^{\rm op}\in  \A\otimes \A^{\rm op}$ normalized by the condition $\sum a_jb_j=1$.  Then the operator $D'=D(\eta(A))$ is equal to the inner fluctuation of $D$ with respect to the algebra $\A\otimes\hat \A$ and the $1$-form $\eta(A\otimes\hat A )$, that is
\begin{equation*}
    D'=D+\sum a_i \hat a_j [D,b_i\hat b_j]
\end{equation*}
$(ii)$~An inner fluctuation of an inner fluctuation of $D$ is still an inner fluctuation of $D$, and more precisely one has, with $A$ and $A'$ normalized elements of $\A\otimes \A^{\rm op}$ as above,
\begin{equation*}
    \left(D(\eta(A))\right)(\eta(A'))=D(\eta(A'A))
\end{equation*}
where the product $A'A$ is taken in the tensor product algebra $\A\otimes \A^{\rm op}$.
\end{prop}
\proof $(i)$~One has, in $\Omega^1(\A\otimes\hat \A)$
\begin{equation*}
    [\delta(b_i),\hat b_j]=\delta(b_i\hat b_j)-b_i\delta(\hat b_j)-\hat b_j \delta(b_i)
\end{equation*}
and thus, using the normalization condition and the commutation of $\A$ with $\hat\A$,
\begin{equation*}
    \sum a_i\hat a_j[\delta(b_i),\hat b_j]=\sum a_i \hat a_j \delta(b_i\hat b_j)
    -\sum a_i\delta(b_i)-\sum \hat a_j\delta(\hat b_j)
\end{equation*}
Applying this with the derivation $[D,.]$ instead of $\delta$ one sees that, in the formula for $D'$, the terms in $A_{(1)}$ and $\hat A_{(1)}$ combine with $A_{(2)}$ to give the required result.

$(ii)$~We let $A=\sum a_j\otimes b_j^{\rm op}$ and $A'=\sum x_s\otimes y_s^{\rm op}$,   both being normalized. We let
\begin{equation*}
    a_{ij}=a_i\hat a_j,\ \ b_{ij}=b_i\hat b_j,\ \
    x_{st}=x_s\hat x_t,\ \ y_{st}=y_s\hat y_t
\end{equation*}
and we have
\begin{equation*}
    D'=D(\eta(A))=D+\sum a_{ij}[D,b_{ij}]
\end{equation*}
and similarly
\begin{equation*}
   D''=D'(\eta(A'))=\left(D(\eta(A))\right)(\eta(A'))=D(\eta(A))+
   \sum x_{st}[D(\eta(A)),y_{st}]
\end{equation*}
which gives
\begin{equation*}
    D''=D+\sum a_{ij}[D,b_{ij}]+\sum x_{st}[D,y_{st}]+
    \sum\sum x_{st}[a_{ij}[D,b_{ij}],y_{st}]
\end{equation*}
Now one has
\begin{equation*}
   x_{st}[a_{ij}[D,b_{ij}],y_{st}]=x_{st}\left(a_{ij}[D,b_{ij}]y_{st}-
   y_{st}a_{ij}[D,b_{ij}]
   \right)
\end{equation*}
and the terms on the right sum up to
\begin{equation*}
  -\sum\sum    x_{st}y_{st}a_{ij}[D,b_{ij}]=-\sum a_{ij}[D,b_{ij}]
\end{equation*}
Moreover one has
\begin{equation*}
    x_{st}a_{ij}[D,b_{ij}]y_{st}=x_{st}a_{ij}[D,b_{ij}y_{st}]-x_{st}a_{ij}b_{ij}[D,y_{st}]
\end{equation*}
and the terms on the right sum up to
\begin{equation*}
    -\sum\sum x_{st}a_{ij}b_{ij}[D,y_{st}]=-\sum x_{st}[D,y_{st}]
\end{equation*}
Thus we have shown that
\begin{equation*}
    D''=D+\sum x_{st}a_{ij}[D,b_{ij}y_{st}]
\end{equation*}
which gives the required result using
\begin{equation*}
    x_{st}a_{ij}=x_s\hat x_ta_i\hat a_j=x_sa_i\hat x_t\hat a_j=x_sa_i\widehat{(x_ta_j)}
\end{equation*}
\begin{equation*}
   b_{ij}y_{st}=b_i\hat b_j y_s\hat y_t=b_iy_s\hat b_j \hat y_t=
   b_iy_s\widehat{(b_jy_t)}
\end{equation*}
and
\begin{equation*}
    \left(\sum x_s\otimes y_s^{\rm op}\right) \left(\sum a_i\otimes b_i^{\rm op}\right)=
    \sum x_sa_i\otimes (b_iy_s)^{\rm op}
\end{equation*}
taking place in the algebra $\A\otimes \A^{\rm op}$.
 \endproof
 Note that the normalization and  self-adjointness conditions are preserved by the product of normalized elements in $\A\otimes \A^{\rm op}$, since
 \begin{equation*}
    \sum x_sa_ib_iy_s=\sum x_sy_s=1
 \end{equation*}
and moreover the following operation is an antilinear automorphism of $\A\otimes \A^{\rm op}$
\begin{equation*}
   \sum a_j\otimes b_j^{\rm op}\mapsto \sum b_j^*\otimes a_j^{*\rm op}
\end{equation*}
while the self-adjointness condition means to be in the fixed points of this automorphism.
It is thus natural to introduce the following semi-group:
\begin{prop}
$(i)$~The self-adjoint normalized elements of $\A\otimes \A^{\rm op}$ form a semi-group $\pert(\A)$ under multiplication.

$(ii)$~The transitivity of inner fluctuations (\ie the fact that inner fluctuations of inner fluctuations are inner fluctuations) corresponds to the semi-group law in the semi-group $\pert(\A)$.

$(iii)$~The semi-group $\pert(\A)$ acts on real spectral triples through the homomorphism $\mu:\pert(\A)\to \pert(\A\otimes \hat \A)$ given by
\begin{equation}\label{mult}
    A\in \A\otimes \A^{\rm op}\mapsto \mu(A)= A\otimes \hat A\in \left(\A\otimes \hat \A\right)\otimes \left(\A\otimes \hat \A\right)^{\rm op}
\end{equation}
\end{prop}
\proof We have shown above that $\pert(\A)$ is a semi-group. Using its action on operators in $\H$  by
$T\mapsto \sum a_iTb_i$ one gets $(ii)$.  Proposition \ref{propinn} gives $(iii)$.  One checks the multiplicativity of the map $\mu$ as follows. Let $A=\sum a_j\otimes b_j^{\rm op}$, $A'=\sum x_s\otimes y_s^{\rm op}$,
$   a_{ij}=a_i\hat a_j,\  b_{ij}=b_i\hat b_j,\
    x_{st}=x_s\hat x_t,\   y_{st}=y_s\hat y_t$
    so that
    \begin{equation*}
        \mu(A)=\sum a_{ij}\otimes b_{ij}^{\rm op}, \ \ \mu(A')=\sum x_{st}\otimes y_{st}^{\rm op}
    \end{equation*}
    Then one has $A'A=\sum x_sa_i\otimes (b_iy_s)^{\rm op}$ and
\begin{equation*}
  \mu(  A'A)=\sum x_sa_i\widehat{(x_ta_j)}\otimes \left(b_iy_s\widehat{(b_jy_t)}\right)^{\rm op}
  =\sum x_{st}a_{ij}\otimes (b_{ij}y_{st})^{\rm op}=\mu(A')\mu(A)
\end{equation*}
which completes the proof of $(iii)$.\endproof

Note that as a subset of $\A\otimes \A^{\rm op}$ the subset $\pert(\A)$ is stable under affine combinations $\alpha A+\beta A'$ for $\alpha,\beta\in \R$ and $\alpha+\beta=1$. The map $\mu$ is quadratic.

To summarize the above discussion we see that the inner fluctuations come from the action of the semi-group $\pert(\A)$ in a way which parallels the action of inner automorphisms and which, for real spectral triples, combines $\A$ with $\hat \A$.
Passing from the ordinary formalism of inner fluctuations for spectral triples to the case of real spectral triples is given by the homomorphism $\mu:\pert(\A)\to \pert(\A\otimes \hat \A)$ on the semi-groups of inner perturbations. The unitary group $\U(\A)$ maps to the semi-group $\pert(\A)$  by the homomorphism
$u\in \U(\A)\mapsto u\otimes (u^*)^{\rm op}\in \pert(\A)$, and this homomorphism is compatible with $\mu$. Moreover the invertible elements of the semi-group $\pert(\A)$ form a group 
 which deserves further investigations.

\section{The spectral $U(1) \times U(2)$-model}

We illustrate the above generalized form of inner fluctuations with the spectral model corresponding to $k=2$ in the classification of irreducible finite geometries of $KO$-dimension $6$ (modulo $8$) performed in \cite{CC07b}. 

The algebra and irreducible Hilbert space representation are:
\begin{align*}
\A &= M_2(\C) \oplus M_2(\C), \\
\H &= \C^2 \otimes \bar{\C}^2 \oplus \C^2 \otimes \bar{\C}^2, 
\end{align*}
acted upon by matrix multiplication from the left ($\C^2$) and from the right ($\bar{\C}^2$). We introduce the following index notation (analogous to \cite{framework}) for vectors in $\H$:
$$
\Psi= \begin{pmatrix}
\psi_A \\ \psi_{A'}
\end{pmatrix}, \qquad \psi_{A'} = \psi^c_A
$$
where $\psi^c_A$ is the conjugate spinor to $\psi_A$. It is acted upon by
both the first and the second copy of $M_{2}\left(  \mathbb{C}\right)$. The index $A$ can take $4$
values and is represented by
$$
A = \alpha I
$$
where the index $\alpha = 1,2$ for the first, and $I=1,2$ for the second copy of $M_2(\C)$. The grading is given by 
$$
\gamma_{\alpha I}^{\beta J} = G_\alpha^\beta \delta_I^J = - \gamma_{\alpha' I'}^{\beta' J'}\qquad \text{with } G_\alpha^\beta = \begin{pmatrix} 1 & 0 \\ 0 & -1 \end{pmatrix}.
$$
This grading breaks the first $M_2(\C) \subset \A$ into $\C_R \oplus \C_L$, where $R$ and $L$ stand for right and left. Thus, the index $\alpha=1$ ($\alpha=2$) corresponds to the $\C_R$ ($\C_L$).
The even subalgebra $\A_\ev = \C_R \oplus \C_L \oplus M_2(\C)$ acts as follows:
\begin{align*}
\pi(\lambda_R, \lambda_L, m) &= \begin{pmatrix}
X_{\alpha}^{\beta}\delta_{I}^{J} & 0\\
0 & \delta_{\alpha^{\prime}}^{\beta^{\prime}}m_{I^{\prime}}^{J^{\prime}}
\end{pmatrix}; \qquad \text{where } X_{\alpha}^{\beta} = \begin{pmatrix}
\lambda_R & 0 \\ 0 & \lambda_L
\end{pmatrix}
\end{align*}
for $(\lambda_R, \lambda_L, m) \in\C_R \oplus \C_L \oplus M_2(\C)$. The real structure with $J^2=1$ and $\gamma J =  - J \gamma$ is given by 
\begin{align*}
J &= \begin{pmatrix} 0 & \delta_{\alpha}^{\beta'} \delta_{I}^{J'} \\ \delta_{\alpha'}^\beta \delta_{I'}^J & 0 \end{pmatrix} \times \text{complex conjugation}.
\end{align*}
This gives for the right action $\pi^\circ(\lambda_R, \lambda_L, m) \equiv J \pi(\lambda_R, \lambda_L, m)^* J^{-1}$:
\begin{align*}
\pi^\circ(\lambda_R, \lambda_L, m) &= \begin{pmatrix}
\delta_{\alpha}^{\beta}m_{I}^{tJ} & 0 \\
0 & X_{\alpha'}^{t\beta'}\delta_{I'}^{J'} 
\end{pmatrix}
\end{align*}
where the superscript $t$ denotes the transpose matrix. This clearly satisfies
the commutation relation
\begin{equation}
[  \pi(a), \pi^\circ(b)]  =0; \qquad (a,b \in \A_\ev).
\end{equation}

Let us now analyze the first-order condition for a Dirac operator of the following form
$$
D = \begin{pmatrix}
D_{A}^{\quad B} & D_{A}^{\quad B^{^{\prime}}}\\
D_{A^{^{\prime}}}^{\quad B} & D_{A^{^{\prime}}}^{\quad B^{^{\prime}}}%
\end{pmatrix}
$$
with 
\begin{align*}
D_{\alpha I}^{\beta J} &:= \left( \delta_{\alpha}^1 \delta^{\beta}_2 k_x 
+ \delta_{\alpha}^2 \delta^{\beta}_1 k_x^* \right)\delta_I^J; \qquad D_{\alpha' I'}^{\beta' J'} = \bar{D_{\alpha I}^{\beta J}},  \\
D_{\alpha' I'}^{\beta J} &:= \delta_{\alpha'}^1 \delta^{\beta'}_1 \delta_{I'}^1 \delta^{J}_1 k_y ; \qquad\qquad\qquad D_{\alpha I}^{\beta' J'} = \bar{D_{\alpha I}^{\beta' J'}}. 
\end{align*}

Due to the presence of the off-diagonal term involving $k_y$, the above spectral triple  $(\A_\ev,\H,D;J)$ does not satisfy the first-order condition:
$$
[[D,\pi(a)], \pi^\circ(b)] = 0.
$$
Instead we have the following result. 
\begin{prop}
The largest (even) subalgebra $\A_F \subset \A$ for which the first-order condition holds (for the above $\H, D$ and $J$) is given by 
$$
\A_F = \left \{ \left( \lambda_R, \lambda_L, \begin{pmatrix} \lambda_R & 0 \\ 0 & \mu \end{pmatrix} \right): (\lambda_R, \lambda_L, \mu) \in \C_R \oplus \C_L \oplus \C \right\} \subset \C_R \oplus \C_L \oplus M_2(\C). 
$$
\end{prop}
\proof
We compute that
$[[D,\pi(\lambda_R,\lambda_L,m)],\pi^\circ(\lambda_R',\lambda_L' ,m') ] =0$ amounts to the vanishing
\begin{gather*}
 \begin{pmatrix} (\lambda_R' -m_{11}') (m_{11}-\lambda_R ) & (\lambda_R' -m_{11}' )m_{12}  \\ -m_{12}' (m_{11}-\lambda_R)  & m_{12}' m_{12}  \end{pmatrix}\bar{ k_y} =0,\\
\begin{pmatrix} (\lambda_R-m_{11})(m_{11}'-\lambda_R') & (\lambda_R -m_{11})m_{21}'  \\ m_{21} (m_{11}' -\lambda_R') & m_{12}' m_{12} \end{pmatrix}k_y=0.
\end{gather*}
The solution set gives the subalgebra $\A_F$.
\endproof

In the next section we will see that the algebra $\A$ can spontaneously break to $\A_F$, using the spectral action on the generalized inner fluctuations defined before for spectral triples that fail on the first-order condition.
As a preparation, we first compute the first-order inner fluctuations $A_{(1)}$ and $JA_{(1)} J^{-1}$, as well as the non-linear term $A_{(2)}$.

\begin{prop}
The inner fluctuated Dirac operator $D'(A)$ is parametrized by three complex scalar fields $\phi, \sigma_1,\sigma_2$ entering in $A_{(1)} \in \Omega^1_D(\A)$ and $A_{(2)}$:
$$
D'(A) =D+A_{(1)}+J A_{(1)}J^{-1} + A_{(2)} \equiv 
\begin{pmatrix}
D'(A)_{\alpha I}^{\beta J} & D'(A)_{\alpha I }^{\beta' J' }\\
D'(A)_{\alpha' I' }^{\beta J} & D'(A)_{\alpha ' I' }^{\beta' J' }%
\end{pmatrix}
$$
where
\begin{align*}
D'(A)_{\alpha I}^{\beta J} &:= \left( \delta_{\alpha}^1 \delta^{\beta}_2 k_x (1+\phi) 
+ \delta_{\alpha}^2 \delta^{\beta}_1 k_x^* (1+ \phi^*) \right)\delta_I^J,\\
D'(A)_{\alpha' I'}^{\beta J} &:= \delta_{\alpha'}^1 \delta^{\beta}_1
(\delta_{I'}^1  + \sigma_{I'} )(\delta^{J}_1+ \sigma^J) k_y.
\end{align*}
\end{prop}
\proof
We parametrize $a_j = (\lambda_{Rj}', \lambda_{Lj}' , m_j')$ and $b_j = (\lambda_{Rj}, \lambda_{Lj} , m_j)$ and compute
$$
A_{(1)} = \sum_j a_j[D,b_j].
$$
If we write $A_{(1)}^* = A_{(1)}$ as
\begin{align*}
A_{(1)}&=\sum_j a_j[D,b_j] = \begin{pmatrix} 
(A_{(1)})_{\alpha I}^{\beta J}  &(A_{(1)})_{\alpha I}^{\beta' J'}\\ 
(A_{(1)})_{\alpha' I'}^{\beta J}  &(A_{(1)})_{\alpha' I'}^{\beta' J'}\\
\end{pmatrix}
\intertext{we compute that $(A_{(1)})_{\alpha' I'}^{\beta' J'}=0$, while}
(A_{(1)})_{1I}^{2J} &=  \sum k_x \lambda_{R}' \left(  \lambda_{L} - \lambda_{R}   \right)\delta_{J}^I \equiv  \bar{(A_{(1)})_{2 J}^{1 I}},\\
%(A_{(1)})_{1 I}^{2 J} &= \sum k_x^* \lambda_{L}' \left(   \lambda_{R} - \lambda_{L}  \right) \delta_{J}^I\equiv (A_{(1)})_{2 J}^{1 I}^*;\\
(A_{(1)})_{\alpha' I'}^{\beta J} &= \sum k_y \delta_{\alpha'}^1 \delta^{\beta}_1 {m'}_{I'}^{K'} \left(\delta_{K'}^1 \delta^{J}_1  \lambda_{R} - m_{K'}^{L'} \delta_{L'}^1 \delta^{J}_1 \right) \equiv \bar{(A_{(1)})_{\alpha I}^{\beta' J'}} .
%&= \sum k_y \delta_{\alpha'}^2 \delta^{\beta}_2 \left( m'_{I'}^{1}\delta^{J}_1  \lambda_{R} - m'_{I'}^{K'} m_{K'}^{1} \delta^{J}_1 \right) 
%\\(A_{(1)})_{\alpha I}^{\beta' J'} &= \sum k_y^* \delta_{\alpha}^2 \delta^{\beta'}_2  \delta_{I'}^1 \delta^{J}_1 \lambda_{R}' \left(  m_{I}^{J'} - \lambda_{R} \delta_I^{J'} \right).
\end{align*}
where for notational clarity we have dropped the index $j$. We parametrize this by complex fields $\phi, \sigma_1, \sigma_2$ as follows:
\begin{align*}
\phi &= \sum \lambda_{R}' (\lambda_{L} - \lambda_{R}), %\equiv \sum_j \bar{\lambda_{Lj}' (\lambda_{Rj} - \lambda_{Lj})}
\\
\sigma_1 &=\sum \left( {m'}_{1}^{1}  \lambda_{R} - {m'}_{1}^{K'} m_{K'}^{1} \right), \\
\sigma_2 &=\sum \left( {m'}_{2}^{1}  \lambda_{R} - {m'}_{2}^{K'} m_{K'}^{1} \right).
\end{align*}
Summarizing:
\begin{align*}
(A_{(1)})_{\alpha I}^{\beta J} &=  k_x \delta_\alpha^1 \delta^\beta_2 \delta_{J}^I \phi + k_x^* \delta_\alpha^2 \delta^\beta_1 \delta_{J}^I \phi^* ,
\\ 
(A_{(1)})_{\alpha' I'}^{\beta J} &=  k_y \delta_{\alpha'}^1 \delta^{\beta}_1 \sigma_{I'} \delta^J_1
= \bar{(A_{(1)})^{\alpha' I'}_{\beta J}},
\end{align*}
and $(A_{(1)})_{\alpha' I'}^{\beta' J'}=0$.

From these expressions and the form of $J$, it follows immediately that
\begin{align*}
(J A_{(1)} J^{-1} )_{\alpha' I'}^{\beta' J'} &= \bar{(A_{(1)})_{\alpha I}^{\beta J}}\\
&=k_x^* \delta_\alpha^1 \delta^\beta_2 \delta_{J}^I \phi^* + k_x \delta_\alpha^2 \delta^\beta_1 \delta_{J}^I \phi,\\
(J A_{(1)} J^{-1})_{\alpha' I'}^{\beta J} &= \bar{(A_{(1)} )_{\alpha I}^{\beta' J'}} \\
&= k_y \delta_{\alpha'}^1 \delta^{\beta}_1 \sigma^J \delta_{I'}^1.
\end{align*}
Next, we determine $A_{(2)}$, which is
\begin{align*}
&A_{(2)} = \sum_j a_j [JA_{(1)} J^{-1},b_j] = 
\begin{pmatrix}
(A_{(2)})_{\alpha I}^{\beta J}  &(A_{(2)})_{\alpha I}^{\beta' J'}\\ 
(A_{(2)})_{\alpha' I'}^{\beta J}  &(A_{(2)})_{\alpha' I'}^{\beta' J'}\\
\end{pmatrix}.
\end{align*}
It follows that $(A_{(2)})_{\alpha I}^{\beta J}=0 = (A_{(2)})_{\alpha' I'}^{\beta' J'}$. On the other hand, we compute in terms of the same elements $a_j = (\lambda_{Rj}', \lambda_{Lj}' , m_j')$ and $b_j = (\lambda_{Rj}, \lambda_{Lj} , m_j)$ as above that
\begin{align*}
(A_{(2)})_{\alpha' I'}^{\beta J}
&=\sum  \delta_{\alpha'}^1 \delta^{\beta}_1  {m'}_{I'}^{K'} \left( \delta_{K'}^1 \sigma^{J}  \lambda_{R} - m_{K'}^{L'}  \delta_{L'}^1 \sigma^J \right).
\end{align*}
One readily checks that the components of $a_j$ and $b_j$ enter in $A_{(2)}$ in precisely the same combinations as before to form the fields $\sigma_1$ and $\sigma_2$. In fact, we simply have
$$
(A_{(2)})_{\alpha' I'}^{\beta J} = k_y \delta_{\alpha'}^1 \delta^{\beta}_1 \sigma_{I'} \sigma^J \equiv \bar{(A_{(2)})^{\alpha' I'}_{\beta J}}.
$$
The result then follows by combining $(D+A_{(1)})_{\alpha I}^{\beta J}$, $(D+JA_{(1)}J^{-1})_{\alpha' I'}^{\beta' J'}$ and finally
\begin{align*}
\left(D+ A_{(1)} +JA_{(1)} J^{-1}+A_{(2)} \right)_{\alpha' I'}^{\beta J} 
&= \delta_{\alpha'}^1 \delta^{\beta'}_1  k_y
\left( \delta_{I'}^1 \delta^{J}_1 +
 \sigma_{I'} \delta^J_1 + \sigma^J \delta_{I'}^1 + \sigma_{I'} \sigma^J \right)\\
&=\delta_{\alpha'}^1 \delta^{\beta'}_1  k_y
( \delta_{I'}^1 +
 \sigma_{I'} )(\delta^{J}_1+ \sigma^J ).
\end{align*}
\endproof

We conclude that the additional term $A_{(2)}$ in the inner fluctuations that is due to the failure of the first-order condition actually does not generate new scalar fields but is parametrized by the same fields that enter in $A_{(1)}$. Even more, the inner fluctuations are of rank 1, and we are dealing with composite, rather than fundamental ``Higgs'' fields. This is quite convenient for the computation of the spectral action, as we will see next.

\subsection{The scalar potential}
Recall that the spectral action \cite{mc2} gives rise to a potential
\begin{align*}
V(\phi, \sigma_1, \sigma_2) &= 
-\frac{f_2}{ 2\pi^2 }\Lambda^2 \tr_\H (D'(A))^2  + \frac{f_0}{8 \pi^2} \tr_\H (D'(A))^4.
\end{align*}
In terms of the fields $\phi, \sigma_1$ and $\sigma_2$ this reads:
\begin{align*}
V(\phi,\sigma_1,\sigma_2)&= -\frac{f_2}{ \pi^2 }\Lambda^2 \left( 4 |k_x|^2 |\phi|^2 + |k_y|^2 (|1+\sigma_1|^2 +  |\sigma_2|^2 )^ 2\right) \\
 &\quad+ \frac{f_0}{4 \pi^2} \bigg( 4 |k_x|^4 | \phi|^4   + 4 |k_x|^2 |k_y|^2 |\phi|^2 (|1+\sigma_1|^2 +  |\sigma_2|^2 )^ 2\\
&\qquad\qquad+ |k_y|^4(|1+\sigma_1|^2 +  |\sigma_2|^2 )^4
\bigg)
\end{align*}
This follows from the explicit form of $D'(A)$ given above and the relation
$$
\tr \left( (v v^ t)^* (v v^ t)\right)^n = |v|^{4n}
$$
which holds for any vector $v$ and $n \geq 0$.

\begin{figure}
\begin{center}
\includegraphics[scale=.6]{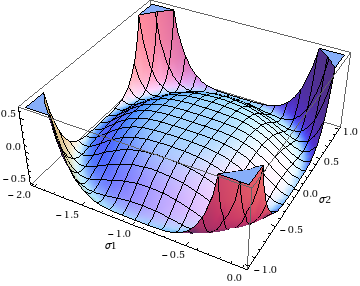}
\end{center}
\caption{The potential $V(\phi=0,\sigma_1,\sigma_2)$.}
\label{fig:potentialsigma1sigma2}
\end{figure}

\begin{prop}
The potential $V(\phi=0, \sigma_1,\sigma_2)$ has a local minimum at $(\sigma_1,\sigma_2) = (-1+\sqrt{w},0)$ with $w=\sqrt{2 f_2\Lambda^2/(f_0 |k_y|^2)}$ and this point spontaneously breaks the symmetry group $\U(\A_\ev)$ to $\U(\A_F)$. 

In fact, including also the field $\phi$, the Hessian matrix of $V$ with respect to $\phi, \sigma_1, \sigma_2$ at $(0,-1+\sqrt{w},0)$ is given by
$$
\textup{Hess}(V) = \frac{f_0 |k_y|^4}{ \pi^2} \begin{pmatrix} -2w^2  & 0 & 0 \\ 0 &8 w^{3} & 0 \\ 0 & 0 &  0 \end{pmatrix}
$$
Thus, upon including the field $\phi$, the point $(\phi,\sigma_1,\sigma_2) = (0,-1+\sqrt w,0)$ is a critical point of $V(\phi, \sigma_1,\sigma_2)$, with the only negative second-derivative in the direction of the $\phi$-field. 
\end{prop}
\proof
The potential is of the following form
$$
V(\phi=0, \sigma_1,\sigma_2) = -\frac{f_2}{ \pi^2 }\Lambda^2 |k_y|^2 (|1+\sigma_1|^2 +  |\sigma_2|^2 )^ 2  + \frac{f_0}{4 \pi^2} |k_y|^4(|1+\sigma_1|^2 +  |\sigma_2|^2 )^4
$$
with minima at $|1+ \sigma_1|^2 + |\sigma_2|^2 = \sqrt{2 f_2 \Lambda^2 /f_0|k_y|^2}$, see also Figure \ref{fig:potentialsigma1sigma2}. In particular, $(\sigma_1,\sigma_2) = (-1 +\sqrt{w},0)$ is one of those minima.

Note that at this point the only non-zero entry in $D' (A)$ is given by $1+2 \sigma_1 + \sigma_2^ 2 \equiv w$. Then, the gauge transformation $D' \mapsto U D' U^*$ with the unitary $U = u JuJ^{-1}$ and $u \in \U(\A_\ev) = U(1) \times U(1) \times U(2)$ implies that this minimum transforms as
$$
D'(A)_{1I'}^{1J} =
\begin{pmatrix} w & 0 \\ 0 & 0 \end{pmatrix}\mapsto
u \bar u_R^2 \begin{pmatrix} w & 0 \\ 0 & 0 \end{pmatrix}  u^t
$$
where $(u_R , u) \in U(1) \times U(2)$. This implies that the only such matrices that leave the minimum invariant are given by
$$
u_R \in U(1), \qquad u = \begin{pmatrix} u_R & 0 \\ 0 & \mu \end{pmatrix}  \quad \text{with } \mu \in U(1).
$$
Since $u_L \in U(1)$ acts trivially, this reduces $\U(\A_\ev) = U(1) \times U(1) \times U(2)$ to $\U(\A_F) = U(1) \times U(1) \times U(1)$. 
\endproof

This mechanism also generates mass terms for the gauge fields corresponding to the coset  $U(1) \times U(1) \times U(2)/U(1) \times U(1) \times U(1)$. As usual, these come from the terms in the spectral action of the form
$$
\frac{f_0}{8 \pi^2} \tr \left|\nabla_\mu (D'(A)) \right|^2
$$
with the gauge fields minimally coupled to the scalar fields $\phi, \sigma_1$ and $\sigma_2$ entering in $D'(A)$. We will not digress on the computational details, and leave a complete treatment for the separate paper \cite{ncgps}, addressing the full ($k=4$) model.

\begin{figure}
\begin{center}
\includegraphics[scale=.5]{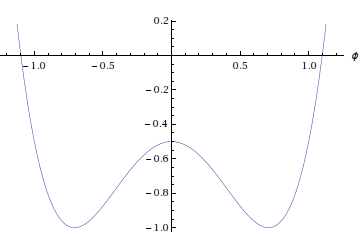}
\end{center}
\caption{The potential $V(\phi,\sigma_1=-1+\sqrt{w},\sigma_2=0)$.}
\label{fig:potentialphi}
\end{figure}

After the fields $(\sigma_1, \sigma_2)$ have reached their vevs $(-1+\sqrt{w},0)$, there is a remaining potential for the $\phi$-field:
$$
V(\phi) = -\frac{2 f_2}{ \pi^2 }\Lambda^2 |k_x|^2|\phi|^2 
 + \frac{f_0}{ \pi^2} |k_x|^4|\phi|^4.
$$
Selecting one of the minima of $V(\phi)$ spontaneously breaks the symmetry further from $\U(\A_F)=U(1) \times U(1) \times U(1)$ to $U(1) \times U(1)$, and generates mass terms for the $L-R$ abelian gauge field. % $B'_\mu :=B_{L\mu}-B_{R\mu}$. 
Again, we leave all computational details for the full model.

\end{document}